\RequirePackage{fix-cm}
\documentclass[smallcondensed]{svjour3}     
\smartqed  
\usepackage{graphicx}
\usepackage{amssymb}

\begin{document}

\title{Parmenides reloaded}



\author{Gustavo E. Romero}


\institute{Instituto Argentino de Radioastronom{\'{i}}a (IAR, CCT La Plata, CONICET) \at
              C.C. No. 5, 1894, Villa Elisa, Buenos Aires, Argentina. \\
              Tel.: +54-221-482-4903\\
              Fax: +54-221-425-4909\\
              \email{romero@iar-conicet.gov.ar}}           

\maketitle

\begin{abstract}
 I argue for a four dimensional, non-dynamical view of space-time, where becoming is not an intrinsic property of reality. This view has many features in common with the Parmenidean conception of the universe. I discuss some recent objections to this position and I offer a comparison of the Parmenidean space-time with an interpretation of Heraclitus' thought that presents no major antagonism. 

\end{abstract}


\section{Theory and reality}

The aim of physical theory is to represent reality (e.g. Bunge 1967). A basic assumption of science is that there are things in the world, and that things have properties (this is an ontological hypothesis). Properties can be represented by mathematical functions and other abstract objects invented by human beings according to self-consistent rules (e.g. Bunge 2006). The value of the functions and the structure of the mathematical objects of the theory are restricted by equations and mathematical conditions that represent physical laws, i.e. restrictions to the possible space of states of things. When the properties of things change, we say that there is an {\it event}. An event is specified by an ordered pair of states of a thing (see Bunge 1977 and the Appendix below). Each state is characterized by a collection of values of state functions. Of course, the characterization of a thing is not unique.  A specific model of a thing depends on those aspects of reality that the theory concerns about. The succession of events (or processes) that occur to a thing forms its {\it history}.

Any physical theory refers to some kind of concrete entities. The existence of these entities is assumed by the theory. If the theory is successful, we gain confidence on the existence of the entities. If the theory fails, we can consider the postulated entities just as exploratory hypothesis that may be abandoned. For instance, the standard model of particle physics assumes the existence of different kind of {\em quantons}: leptons, quarks, and interaction bosons. The existence of $W^{\pm}$ and $Z^{0}$ bosons was a working hypothesis of the electroweak interaction model till the particles were discovered in experiments at the end of the 1970s. Now, they are considered as real as photons. Tachyons, on the contrary, are currently discredited as constituents of the world, because of both theoretical reasons and lack of observational evidence. Other particles, like the Higgs boson, remain hypothetical but plausible so far.

The kind of objects that physical theories assume as elements of the world can change as our knowledge of the world evolves. From plants, animals, and planets to elementary particles and quark stars, our view of the universe can change, changing what we think there is, how things are, and how they relate.              

\section{The manifold model of space-time}

General relativity is the theory of space, time and gravitation formulated by Albert Einstein in 1915 (Einstein 1916). It is an extraordinarily successful theory that has passed many test, both in the weak and strong field limits. It is a highly complex theory where the gravitational field is described through the curvature of space-time. The field equations are ten non-linear differential equations in the coefficients of the metric tensor of space-time. The theory reaches its maximum predictive power when expressed independently of coordinates in the language of abstract differential geometry. In this formulation, known as the {\sl manifold model of space-time} (e.g. Hawking \& Ellis 1973, Joshi 1993), the appearance of coordinate singularities as that in Schwarzschild's solution can be avoided. The theory can be applied from compact objects with strong gravity to the whole universe. 

The basic concept of this formulation of general relativity is the concept of {\em space-time}, introduced by Hermann Minkowski in 1908 (Minkowski 1909). Space-time can be defined as {\sl the ontological sum\footnote{I write {\it ontological sum} and not `mereological aggregation' because I do not consider space-time as a thing nor an individual, but as an emergent property of all changing things. See Perez Bergliaffa et al. (1998).} of all events of all things}. This is not a mere set, which is a mathematical object (i.e. a fiction), but an emergent relational property of all things. Everything that has happened, everything that happens, everything that will happen, is just an element of space-time. 

As it is the case with every physical property, we can represent space-time with some mathematical structure, in order to describe it. The mathematical structure and the property represented should not be confused: the correspondence is never perfect, it always remain tentative. The manifold model of space-time adopts the following mathematical structure:\\

{\sl Space-time can be represented by a $C^{\infty}$ differentiable, 4-dimensional, real manifold.}\\

A real 4-D manifold is a set that can be covered completely by subsets whose elements are in a one-to-one correspondence with subsets of $\Re^{4}$. Every event is represented by a point of the manifold (the converse is not necessary true). Each element of the manifold {\it represents} an event. We adopt 4 dimensions because it seems enough to give 4 real numbers to localize an event (i.e. to provide a minimum characterization). We can always provide a set of 4 real numbers for every event, and this can be done independently of the intrinsic geometry of the manifold. If there are more than a single characterization of an event, we can always find a transformation law between the different coordinate systems. This is a basic property of manifolds. 

If we want to calculate distances between two events, we need more structure on the manifold: we need a geometric structure. We can get it introducing a metric tensor field $g_{ab}$ to determine distances. The infinitesimal separation between two elements of the manifold, which represent two events of space-time, is given by:

\begin{equation}
ds^{2}=g_{ab} dx^{a} dx^{b}.
\end{equation}

Space-time, then, is fully represented by an order pair $(M, g)$, where $M$ is the manifold and $g$ is the metric tensor field. In general relativity the metric of space-time is determined by the energy-momentum of the physical systems through the Einstein's field equations. The metric itself then represents the gravitational potential and its derivatives determine the equations of motion through the affine connection of the manifold. 

As it is the case of any other physical theory, the manifold model of space-time assumes some entities that are represented mathematically. The basic assumption here is the existence of what is represented by the points of the manifold: the totality of events, the changes of all things (and, hence, of such things, since there are no changes without changing things). 

Since the manifold is 4 dimensional, a process, or even the whole history of a 3-dimensional thing, can be represented by a 4-dimensional object (e.g. Heller 1990, Balashov 2010). Although human experience of change can be used to inspire the concept of manifold, once it is adopted, we can describe space-time from a 4-dimensional point of view, where there is no global change. Change of space-time would require an extra dimension not included in space-time. This, in turn, would imply that space-time is a thing with an emergent relational property that should be measured by the extra dimension or `meta-time'. There is no physical reason to introduce such an ontology. And if someone is willing to pay the price to do it, an infinite regress follows immediately, since the 5D `super space-time' might change requiring more extra dimensions ontological inflation would turn the price unaffordable.  

Strings of changes and irreversible processes of physical things are described by asymmetries, intrinsic features, of space-time. Dynamics is the result of comparing different slices of space-time. The `present' is not a moving thing. It is just a concept, a class of events. All this conforms the so-called {\it block universe} ontology (e.g. Smart 1963, Balashov 2010). This view was also expressed, rather poetically, by Hermann Weyl (Weyl 1949):

\begin{quotation}
	The objective world simply is, it does not happen. Only to the gaze of my consciousness, crawling upward along the life line of my body, does a section of this world come to life as a fleeting image in space which continuously  changes in time. 
\end{quotation}

\section{The Parmenidean universe}

Parmenides was born and lived in Elea, a town on the west cost of southern Italy, from the end of the 6th century to the mid 5th century BC. He wrote a poem in hexameters entitled {\em On What Is}. Almost the whole first part of the poem and fragments of the second part have survived thanks to Simplicius, who copied part of the text in the 6th century AD into his commentary of Aristotle's {\em Physics}. 

The first part of the poem is called {\em The Way of Truth}. This piece contains the first known example of a deductive system applied to physical reality. Parmenides was not content just with giving his view of the world. He supported his interpretation of world by logical deduction from what he considered self-evident premises. He stated that there is no change, no becoming, no coming to be. Reality turns out to be unchanging, eternal, motionless, perfect, and single. There is just one thing: the World. His monism is absolute. What we think is a changing world is only the result of illusion and deception. 

The premises of Parmenides' argument can be written as:

\begin{itemize}
\item What is, is.
\item What is not, is not.
\end{itemize}

Then, nothing can come to be from what is not, because `what is not' is not something. {\it Creatio ex nihilo} is nonsense. Change is impossible since, for Parmenides, change is the occupation of empty space, but there cannot be `empty space'. Reality must then be an unchanging block.  

Many centuries later, with the advent of field theories it became clear that change can occur even in a full universe: change does not require empty space. A perturbation in a field that fills the whole universe is a change. 

The concept of change is central to the manifold model of space-time. But once the geometry of the manifold is determined by a tensor field representing the distribution of energy and momentum, its structure is fixed. The universe is represented by the triplet $(M,\;g,\;T)$, where $T$ is the tensor field that represents the properties (energy and momentum) of things.  Points of the manifold represent events, but there is no event or change affecting the space-time as a whole. The four dimensional space-time, mathematically represented by the manifold, is unchanging, eternal, motionless, single, just as the Parmenidean universe. What we call irreversible processes are described by asymmetries in the manifold. The objects that populate the universe are 4-dimensional. They have `temporal parts', as well as spatial parts. In this way, the child I was, is just a part of a larger being, I, that is 4-dimensional. What we call `birth' and `death' are just temporal boundaries of such a being. Change appears only when we consider 3-dimensional slices of 4-dimensional objects. In words of Max Tegmark:

\begin{quotation}
Time is the fourth dimension. The passage of time is an illusion. We have this illusion of a changing, three-dimensional world, even though nothing changes in the four dimensional union of space and time of Einstein's relativity theory. If life were a movie, physical reality would be the entire DVD: Future and past frames exist just as much as the present one\footnote{From an interview by Eaves (2008).}.
\end{quotation}

It seems not to be unfair to call this interpretation of space-time a Parmenidean view of the world. Parmenides, we might say, is back with a vengeance, in 4 dimensions.

\section{Objections}

Recently, Mario Bunge has forcibly criticized the interpretation of the manifold model of space-time outlined above (Bunge 2011). The core of his argument is the following:

\begin{quotation}

If points in a spacetime grid are identified with events,
instead of being said to represent possible events, becoming vanishes.
But this is absurd: you are still alive, your great-grand children
are not yet born, the next economic bust is yet to come, the Sun has
not yet imploded, and so on. Novelties occur objectively all the time, even if the origin of time is conventional. 
Since the neo-Parmenidean conclusion is utterly false, its premise
must be false as well. What was the premise? That 
spacetime points $=$ events
instead of spacetime points represent (point) events. That is, the trouble in question is a semantic fallacy: that of identifying the map with the territory, the portrait with the subject, the wiring diagram with the network, the model with its referent.

\end{quotation}

I maintain, however, that there is no semantic fallacy here. As it is clear from the definitions given above, space-time has no `points'. Space-time has been defined as an aggregation of {\it events}. It is an emergent relational property of all changing things. The manifold, which is certainly a mathematical concept, {\it represents} space-time, and the elements of the manifold represent events. I do not subscribe that space-time is a {\it thing}, as the sustantivalists do (e.g. Nerlich 1994). I hold that space-time is an emergent relational property of all material things. This is the very same position maintained by Bunge (1977) and developed by Perez-Bergliaffa et al. (1998), among others. The `portrait' is the manifold model, the `subject', space-time, and space-time emerges from changing things. So, the ontological assumption of the manifold model of space-time is that there are changing things, something that Bunge probably will not deny. But space-time itself cannot change, unless we accept a multidimensional time, with the extra dimensions not included in space-time (see arguments against multidimensional time in Bunge 1958). 

The emergence of space-time from changing basic things is also essential to the foundations of background-independent quantum gravity (e.g. Rovelli 2004). If things relate discontinuously, then space-time itself should display quantum features.

Another argument used by Bunge is:

\begin{quotation}

...the notion that time is just one more geometric dimension, on a par with the other three (or seven), is false as well, as shown by the privileged role it holds in the equations of motion. For example, Hamilton's equation $dp/dt = - \partial H/\partial x$ has no spatial counterpart. Likewise, the boundary conditions, so important in continuum mechanics and in quantum mechanics, have no temporal counterparts.
\end{quotation}
    
These statements are based on an incorrect interpretation of the manifold model. Time {\it is not} on a par with the other dimensions, since the space-time metric is represented by a tensor field of trace $-2$, not 4. The manifold is locally Lorentzian, not Euclidean, hence it represents correctly the distinctive role of time. A mere `spatialization' of time is inconsistent with our current knowledge of nature. Inconsistencies, moreover, can be found in the presentism advocated by Bunge: in a universe with a finite and constant velocity for the propagation of interactions, simultaneity is not absolute as in a Newtonian space (Einstein 1905) and past and future of events not causally connected are relative to a reference system. This does not mean that some events exist with respect to one system and not to other. The constraints on what can be known from some physical reference frame are epistemological, not ontological. Existence is invariant under general coordinate transformation. The events that we call `future' are as real as those we call `past' (see, e.g., Putnam 1967).

Regarding boundary conditions, their temporal counterparts are the so-called initial conditions. Actual boundary conditions in space-time should be fixed in 4 dimensions in order to make predictions. We should not confuse the predictive power of our theories, with ontological determinism. The latter is a metaphysical doctrine: the doctrine that all events exist, independently of our possibility of knowing or predicting them. The manifold model of space-time is ontologically deterministic, although it is compatible with epistemic indeterminacy.  

It might be the case that the manifold model of space-time be ultimately incorrect. All representations of reality are just imperfect approximations, but the problems of the model are not semantic. They are more likely related to the applicability of the manifold concept to represent events at the Planck scale.

\section{Heraclitus' river}

It is usual to oppose Parmenides' ideas of a changeless reality to Heraclitus' view that ``all things flow''. Such a view, however, is not based on the extant fragments, but on Plato's interpretation, presented in his {\em Cratylus} (DK 22A6\footnote{The notation refers to the doxography in H. Diels and W. Kranz, {\em Die Fragmente der Vorsokratiker}, 6th ed., Berlin, 1951.}):

\begin{quotation}
All things move and nothing remains, and likening existing things to the flow of a river he says that you could not step twice into the same river.
\end{quotation}

The origin of this seems to be in Heraclitus' fragment DK 22B12:

\begin{quotation}
Upon those who step into the same rivers, different and different waters flow. 
\end{quotation}

In the {\em Theaetetus}, Plato goes further and attributes to Heraclitus the view that all things are always changing in all respects. As pointed out by McKirahan (1994), Plato is likely considering not Heraclitus' thought, but some radical elaborations made much later by Heracliteans or perhaps even by Cratylus.  Heraclitus' fragments seem to emphasize stability through change, and not a changing KOSMOS ($\kappa \acute{o} \sigma\mu o \varsigma$). Morever, a KOSMOS based only on change without stability or HARMONIA ($\grave{\alpha}\rho\mu o \nu \acute{\iota} \alpha$), is a contradiction in terms, a {\em contradictio in adjecto}: it would be CHAOS ($\chi\acute{\alpha}o\varsigma$).  The most important idea in the fragments is that there is a LOGOS ($\lambda \acute{o}\gamma o \varsigma$) in the KOSMOS, a kind of general principle that applies to everything. If we remain faithful to the extant fragments, we see that Heraclitus states that stability is achieved by continuous change. If a river does not flow, if it does not `contain' change, it is not a river, it is a lake. It is just by changing that the river achieves its stability. The same can be extended to all things. Fragment DK 22B84a:  

\begin{quotation}
By changing it is at rest. 
\end{quotation}
   
Heraclitus, moreover, shares some ontological and epistemological concerns with Parmenides, as shown in DK 22B50 and DK 22B123, respectively:

\begin{quotation}
Listening not to me but to the LOGOS is wise to agree that all things are one. 
\end{quotation}

\begin{quotation}
Nature loves to hide. 
\end{quotation}

I offer the suggestion that the ontological antagonism between Parmenides and Heraclitus usually mentioned by so many authors is the result of a doxographic tradition that has its origin in Plato. There is not much in Parmenides' poem nor in Heraclitus' extant fragments to support a frontal opposition. The KOSMOS (or space-time in a modern view) might be changeless and nonetheless formed by changing things, as Heraclitus' river.  

\section{Conclusion: time does not go by}

I do not support the spatialization of time. I maintain that space-time, an emergent property of all things, cannot change. There is nothing respect to which space-time might change. Irreversible processes are represented by asymmetries in the foliation of the manifold that provides a model for space-time. Space-time can be modeled because is part of the physical reality, like any other relational property. Time does not flow. It cannot flow because it is not a thing. Time does not go by. We do.

\section*{Appendix}

\subsection*{1. Events}

In what follows I shall provide a more rigorous characterization of the fundamental concept of \textit{event}.

The concept of individual is the basic primitive concept of any ontological theory. I follow Bunge (1977) on the basics of the ontological views presented here. Individuals associate themselves with other individuals to yield new individuals. It follows that they satisfy a calculus, and that they are rigorously characterized only through the laws of such a calculus. These laws are set with the aim of reproducing the way real things associate. Specifically, it is postulated that every individual is an element of a set $s$ in such a way that the structure $\textsl{S}=\left\langle s, \circ, \square \right\rangle$ is a \textit{commutative monoid of idempotents}. This is a simple additive semi-group with neutral element.

In the structure \textsl{S}, $s$ is the set of all individuals, the element $\square \in s$ is a fiction called the null individual, and the binary operation $\circ$ is the association of individuals. Although \textsl{S} is a mathematical entity, the elements of $s$ are not, with the only exception of $\square$, which is a fiction introduced to form a calculus. The association of any element of $s$ with $\square$ yields the same element. The following definitions characterize the composition of individuals.

\begin{enumerate}

\item ${x} \in s $ is composed $ \Leftrightarrow  \left(\exists {y}, {z}\right)_{s} \left( {x} ={y} \circ {z} \right) $

\item ${x} \in s $ is simple $ \Leftrightarrow \; \sim \left(\exists {y}, {z}\right)_{s} \left({x} ={y} \circ {z} \right)$

\item $ {x}\subset {y}\ \Leftrightarrow {x} \circ {y} = {y}\ $ (${x}$ is part of ${y}\ \Leftrightarrow {x} \circ {y} = {y} $) 

\item $ \textsl{Comp}({x}) \equiv\{{y}\in s \;|\; {y}\subset {x}\}$ is the composition of ${x}$.\\ 

\end{enumerate}

An individual with its properties make up a thing $X$:
\[
	X=<x,P(x)>
\]	
	
Here $P(x)$ is the collection of properties of the individual $x$. A material thing is an individual with material properties, {\em i.e.} properties that can change (see below) in some respect.

Things are distinguished from abstract individuals because they have a number of properties in addition to their capability of association. These properties can be \textit{intrinsic} \normalfont ($P_i$) or \textit{relational} \normalfont ($P_r$). The intrinsic properties are inherent and they are represented by predicates or unary applications, whereas relational properties depend upon more than a single thing and are represented by $n$-ary predicates, with $n \geq 1$. Examples of intrinsic properties are electric charge and rest mass, whereas velocity of macroscopic bodies and volume are relational properties. Velocity (actually its modulus) is an intrinsic property only in the case of photons and other bosons that move at the speed of light in any reference system.

The {\em state} of a thing $X$ is a set of functions $S(X)$ from a 
domain of reference $M$ (a set that can be enumerable or nondenumerable) to the set of properties ${\cal P}_{X}$. Every function in $S(X)$ represents a property in ${\cal P}_{X}$. The set
of the {\sl physically accessible} states of a thing $X$ is the {\em lawful state space} of
$X$: $S_{\rm L}(X)$. The state of a thing is represented by a point in
$S_{\rm L}(X)$. A change of a thing is an ordered pair of states. Only changing things can be material. Abstract things cannot change since they have only one state (their properties are fixed by definition).

Finally, an {\sl event} is a change of a thing $X$, {\em i.e.} an ordered pair of states:
\[ (s_1, s_2 ) \in E_{\rm L}(X) = S_{\rm L}(X) \times S_{\rm L}(X) \]

The space $E_{\rm L}(X)$ is called the {\em event space} of $X$.

The \textit{Universe} $\cal{U}$ is the composition of all things:  $(\neg\exists X) (X \neg\subset \cal{U})$. Space-time is an emergent property of $\cal{U}$ (see Perez Bergliaffa et al. 1998 for details).

\subsection*{2. Presentism}

Presentism, the doctrine advocated by Bunge, can be defined roughly like the thesis that only the present is real. More precisely (see Crisp 2010):

\begin{quotation}
{\em Presentism}. It is always the case that, for every $x$, $x$ is present. 
\end{quotation}

`Present' seems to refer to a given instant of time. Time is usually represented by a 1D real continuum. Of course, the choice of the origin of coordinates when we adopt a metric for time is conventional. Events signaled by different instants are not. What event is present?. When is present?. The answer seems to be {\em now}. `Now', as I have suggested elsewhere, seems to be a class of events that are related to a given brain state (Romero 2011, see also Gr\"unbaum 1973). If this hypothesis is correct, the `present' is a construction of the brain based on its interaction with a class of changing things that affects it. The `present' is not a thing, that moves from past to future. Every conscious brain process has its own present. 

Some people think of the present as a kind of boundary between what existed (and somehow is vanished) and what does not exist: the future. Things pop out from nothing, exist during an unspecified span, and then vanish forever. This violates what is perhaps the most basic principle of science, a principle introduced by Parmenides: nothing comes out of nothing. Presentism implies that everything comes out of nothing, all the time, and disappears into nothing after an indivisible interval of time. Even Heraclitus, I venture, would be terrified.

Presentism is also incompatible with current physics. Let us consider two events: I type this line (event $e1$), and a supernova explodes in the galaxy M83 ($e2$). These events can be considered simultaneous for some `observers' (i.e. when measured in some reference frame), or related by `$e1$ is earlier than $e2$' in other frame, or `$e1$ is later than $e2$' in yet another frame. The fact is that both events exist, be present or not for some people. Similarly, Parmenides exist in some region of space-time, that covers Elea and part of Ancient Greece between, say, 515 and 450 BC. And for some time, he shared his present with Zeno. I exist beyond the space-time region occupied by Parmenides. We shall never meet. But we both are part of the same space-time. I feel lucky for that.         

\begin{acknowledgements}
I am very grateful to Prof. Mario Bunge for insightful discussions and devastating criticisms. I thank Santiago Perez Bergliaffa, V. Bosch-Ramon, and Daniela P\'erez for valuable comments on the manuscript. I {\it should have been} supported by research grant PIP 0078 from CONICET.

\end{acknowledgements}



\bibliographystyle{aipproc}   


\newpage

\section*{Gustavo E. Romero} Full Professor of Relativistic Astrophysics at the University of La Plata and Chief Researcher of the National Research Council of Argentina. He has published more than 250 papers on astrophysics, gravitation, and the foundation of physics, and 8 books. His main current interest is on black hole physics and ontological problems of space-time theories.   










\end{document}